\documentclass[a4paper,twocolumn,prl]{revtex4}
\usepackage[latin9]{inputenc}
\usepackage{amsmath}
\usepackage{amssymb}
\usepackage{graphicx}
\usepackage{esint}

\makeatletter

\usepackage{color}

\usepackage{dcolumn}\usepackage{bm}\usepackage{epsfig}

\makeatother

\begin{document}

\title{Photonic Simulation of System-Environment Interaction: Non-Markovian
Process and Dynamical Decoupling}

\author{Chang-Ling Zou}
\affiliation{Key Lab of Quantum Information, University of Science and Technology
of China, Hefei 230026, Anhui, China}

\author{Xiang-Dong Chen}
\affiliation{Key Lab of Quantum Information, University of Science and Technology
of China, Hefei 230026, Anhui, China}

\author{Xiao Xiong}
\affiliation{Key Lab of Quantum Information, University of Science and Technology
of China, Hefei 230026, Anhui, China}

\author{Fang-Wen Sun}
\email{fwsun@ustc.edu.cn}
\affiliation{Key Lab of Quantum Information, University of Science and Technology
of China, Hefei 230026, Anhui, China}

\author{Xu-Bo Zou}
\affiliation{Key Lab of Quantum Information, University of Science and Technology
of China, Hefei 230026, Anhui, China}

\author{Zheng-Fu Han}
\affiliation{Key Lab of Quantum Information, University of Science and Technology
of China, Hefei 230026, Anhui, China}

\author{Guang-Can Guo}
\affiliation{Key Lab of Quantum Information, University of Science and Technology
of China, Hefei 230026, Anhui, China}

\date{\today}
\begin{abstract}
The system-environment interaction is simulated by light propagating
in coupled photonic waveguides. The profile of the electromagnetic
field provides intuitive physical insight to study the Markovian and
non-Markovian dynamics of open quantum systems. The transition
from non-Markovian to Markovian process is demonstrated by increasing the size of
environment, as the energy evolution changes from oscillating to an
exponential decay, and the revival period increases. Moreover, the dynamical
decoupling with a sequence of phase modulations is introduced to such a photonic open system to form a band structure in time dimension, where the
energy dissipation can be significantly accelerated or inhibited. It
opens the possibility to tune the dissipation in photonic
system, similar to the dynamic decoupling of spins.

\end{abstract}


\maketitle
\emph{Introduction.- }Quantum systems always inevitably interact with
environments, which can change the quantum
states and lead to energy dissipation and decoherence of the systems
\citep{Breuer,RevModPhys.75.715}. For a deep understanding of quantum
effects and broad application of quantum matters, great
efforts have been made to study the system-environment interaction. Lots of interesting phenomena of open quantum system have
been studied, such as the Zeno and anti-Zeno effects \citep{fischer2001observation,koshino2005quantum},
transition between quantum Markovian and non-Markovian processes \citep{BHLiu,PhysRevLett.101.150402}.
Advanced quantum techniques related to the environment have been developed
and remarkable progress has been achieved in experiments, such as
the dissipation engineering to prepare and control quantum states
with the help of environment \citep{diehl2008quantum,verstraete2009quantum,barreiro2010experimental,krauter2011entanglement},
and dynamical decoupling to preserve the quantum coherence from the
environment noise \citep{PhysRevLett.82.2417,du2009preserving,de2010universal}. However, it is still difficult to fully control the high-dimensional environment, which highly limits 
the understanding, control and application of this system-environment interaction.

Recently, increasing experimental and theoretical efforts
are focused on quantum simulations \citep{buluta2009quantum,bloch2012quantum,aspuru2012photonic},
which was inspired by Feynman's seminal idea
\citep{feynman1982simulating}. Various complex and important physical
phenomena can be studied with quantum simulators with high efficiency,
such as the quantum decoherence \citep{barreiro2011open}, many-body
physics, the mechanism of superconducting and the general relativity
\citep{bloch2012quantum}. Such quantum simulations can provide
different views to study subtle physical processes and reveal new
phenomena. And more important, people can also learn new ideas from
the complementary interdiscipline, and exploit the quantum physics
to design new devices for quantum technology and practical applications.
For example, the photonic simulation of electron spins can be used for
the topological protected delay \citep{hafezi2011robust}, and the
simulation of adiabatic passage in waveguide can be applied to high-efficiency
optical coupler \citep{longhi2009quantum,garanovich2012light}.

In this paper, we simulated the quantum open system by photons in
integrated photonic chip. The underlying physics of the system-environment
interaction are revealed intuitively by simply observing the electromagnetic
field profile. The non-Markovian and Markovian processes of this photonic
open system are studied by varying the size of environment. Due to
the memory effect of finite environment as the energy leaked to environment
will be reflected back by the boundary of environment, the dynamics
of the system is non-Markov and shows non-exponential decay and revival.
While for the reservoir with infinite size, the system shows Markovian
exponential decay. Furthermore, we demonstrated the control of the
system decay by dynamical decoupling. With an external modulation
of the phase of photon in system, the dissipation to environment can
be accelerated or inhibited. Our study provides an intuitive understand
of the system-environment interaction, and the results can also be
used to analyze and reduce the leakage losses of photonic structures.

\begin{figure}
\includegraphics[width=8cm]{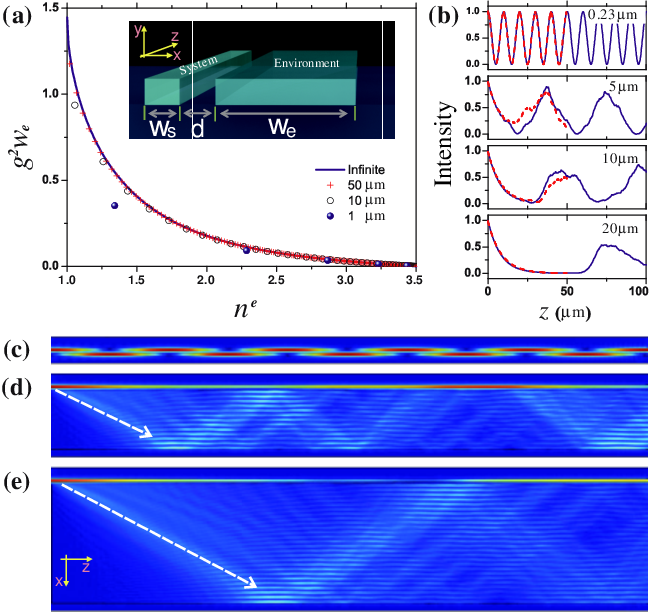} \caption{(Color online) (a) The normalized coupling strength $g^{2}w_{e}$
between system and environment against effect mode index $n^{e}$
for different environment size $w_{e}$. Inset: schematic illustration
of the coupled photonic waveguides simulation to the system-environment
interaction. (b) Typical dynamics of system energy against the interaction
length $z$ for different $w_{e}$, with the solid and dashed lines
are analytical and numerical results. (c)-(e) False-color profiles
of electric field intensity for light loaded in the system waveguide
at $z=0$, with $d=0.15\mu m$ and $w_{e}=0.23,5,10$ $\mu m,$ respectively.}

\label{f1}
\end{figure}

\emph{Model.-} As illustrated by the Inset of Fig. 1(a), the proposed
photonic simulation of the system-environment interaction is composed
of two separated waveguides. Photon can be loaded at \emph{system}
waveguide and travel along it ($z$-axis). The system is open because
its energy can dissipate to the nearby waveguide which acts as \emph{environment}.
The dynamics of photon in single isolated waveguide can be described
by the Helmholtz equations as \citep{yariv1989quantum}
\begin{equation}
\nabla^{2}\psi(\overrightarrow{r})+V(\overrightarrow{r})\psi(\overrightarrow{r})=E\psi(\overrightarrow{r}),
\end{equation}
where $V(\overrightarrow{r})=[\varepsilon(\overrightarrow{r})-1]k^{2}$
and $E=k^{2}$, with the dielectric relative permittivity $\varepsilon(\overrightarrow{r})$ and the wave-number $k=2\pi/\lambda$. The waveguides are uniform along the $z$-axis. Therefore
the $i$-th propagating eigenmode's wave function can be expressed
as $\psi_{i}(\overrightarrow{r})=\varphi_{i}(\overrightarrow{x})e^{in_{i}kz}$,
where $\varphi_{i}(\overrightarrow{x})$ is the field distribution
at the cross section of a waveguide, and $n_{i}$ is the effective
mode index which can be solved from the characteristic equation. For
coupled system and environment waveguides, we can decompose any field
distribution perturbatively as
\begin{equation}
\psi(\overrightarrow{r})=\sum_{i}c_{i}^{s}(z)\psi_{i}^{s}(\overrightarrow{r})+\sum_{j}c_{j}^{e}(z)\psi_{j}^{e}(\overrightarrow{r}),
\end{equation}
where superscript \emph{s} and \emph{e} denote system and environment
waveguides, respectively. Here, $i=1,...,N_{s}$, $j=1,...,N_{e}$,
with $N_{s(e)}$ is the number of modes in the system (environment)
waveguides. $\psi_{i}^{s}(\overrightarrow{r})$ and $\psi_{j}^{e}(\overrightarrow{r})$
are eigenmodes of waveguides, and $c_{i}^{s}(z)$ and $c_{j}^{e}(z)$
are corresponding coefficients. Substituting $V(\overrightarrow{r})=V^{s}(\overrightarrow{r})+V^{e}(\overrightarrow{r})$
and $\psi(\overrightarrow{r})$ to Eq. (1), we can obtain the dynamics
of the modes in each waveguide as
\begin{align}
i\frac{\partial c_{i}^{s(e)}}{\partial z} & =\sum_{l}m_{il}^{s(e)}e^{i(n_{i}^{s(e)}-n_{l}^{s(e)})kz}c_{l}^{s(e)}\nonumber \\
 & +\sum_{j}g_{ij}^{s(e)}e^{i(n_{i}^{s(e)}-n_{j}^{e(s)})kz}c_{j}^{e(s)},
\end{align}
where coupling efficiencies $m_{il}^{s(e)}=\frac{\int(\varphi_{i}^{s(e)})^{\ast}V^{s(e)}\varphi_{l}^{s(e)}d\overrightarrow{x}}{2n_{i}^{s(e)}k}$
and $g_{ij}^{s(e)}=\frac{\int(\varphi_{i}^{s(e)})^{\ast}V^{s(e)}\varphi_{j}^{e(s)}dx}{2n_{i}^{s(e)}k}$,
with all wave functions are normalized by $\int\left\vert \varphi_{i}^{s(e)}\right\vert ^{2}dx=1$.

Considering a thin system waveguide that supports only one guiding
mode ($N_{s}=1$), while the number of modes of the environment waveguide
$N_{e}\geq1$ which depends on its width $w_{e}$. Denoting the system
and environment fields by $a=c_{1}^{s}e^{-in_{1}^{s}kz}$ and $b_{j}=\sqrt{g_{1j}^{s}/g_{1j}^{e}}c_{j}^{e}e^{-in_{j}^{e}kz}$
($j=1,...,N_{e}$) respectively, and replacing the space coordinate
$z$ by time $t$, we can obtain the Hamiltonian which governs the
dynamics of this photonic simulation of the system-environment interaction
as ($\hbar=1$)
\begin{equation}
H=\beta_{0}a^{\dag}a+\sum_{j}\beta_{j}b_{j}^{\dag}b_{j}+\sum_{j}g_{j}(a^{\dag}b_{j}+ab_{j}^{\dag}),\label{Hamiltonian}
\end{equation}
where $\beta_{0}=(n_{1}^{s}+m_{11}^{s})k=n_{0}k$ and $\beta_{j}=(n_{j}^{e}+m_{jj}^{e})k=n_{j}k$
are propagation constants, and $g_{j}=\sqrt{g_{1j}^{s}g_{1j}^{e}}$
is the coupling coefficients. Since the direct coupling between environment
modes is much smaller than other terms, $m_{i,j}^{e}$ ($i\not=j$)
is neglected in following studies.

\emph{Non-Markovian and Markovian Processes.-} The Hamiltonian of
Eq. (\ref{Hamiltonian}) resembles an open system that a harmonic
oscillator ($a$) couples to a collection of oscillators in environment
($b_{j}$). Here, both the coupling coefficients and the environment can be well controlled, which is very promising to simulate the system-environment interaction. With two-dimensional approximation, $\beta_{j}$ and $g_{j}$
can be solved analytically. In the following, we studied the model
with photonic waveguides made of silicon $(n_{d}=3.5)$, which have
been extensively studied in practical experiments. The working wavelength
$(\lambda)$ is $1550\mathrm{nm}$, and the width of system waveguide
is fixed to $w_{s}=0.23\mathrm{\mu m}$ in the single mode regime.

The normalized coupling strength ($g^{2}w_{e}$) between system and
environment against mode index $n^{e}=\beta_{j}/k$ is shown in Fig.
1(a). The continuum of finite size environment is quantized to discrete
modes, whose density and number $N_{e}$ increases with $w_{e}$.
The behaviors of $g^{2}w_{e}$ converges to a line with infinite $N_{e}$
when $w_{e}$=$\infty$, which corresponds to system-reservoir interaction.
With these parameters, the dynamics of the system and environment
can be solved according to Eq. (\ref{Hamiltonian}). In Fig. 1(b),
the evolutions of energy in system waveguide against $z$ for different
$w_{e}$ are calculated, with light loaded in the system waveguide
at $z=0$. For a comparison, the fields in system are also simulated
numerically by finite element method. The results of our analytical
method agree very well with numerical results, with small discrepancies
originating from the slow varying and weak coupling approximations
used in our analytical expressions. For small environment ($w_{e}=0.23\mu m$
and $5\mu m$), the dynamics of the system shows periodic oscillation.
When the environment size increases, the dynamics changes: the system
energy shows exponential decay at first and revives after a distance.

\begin{figure}
\includegraphics[width=8cm]{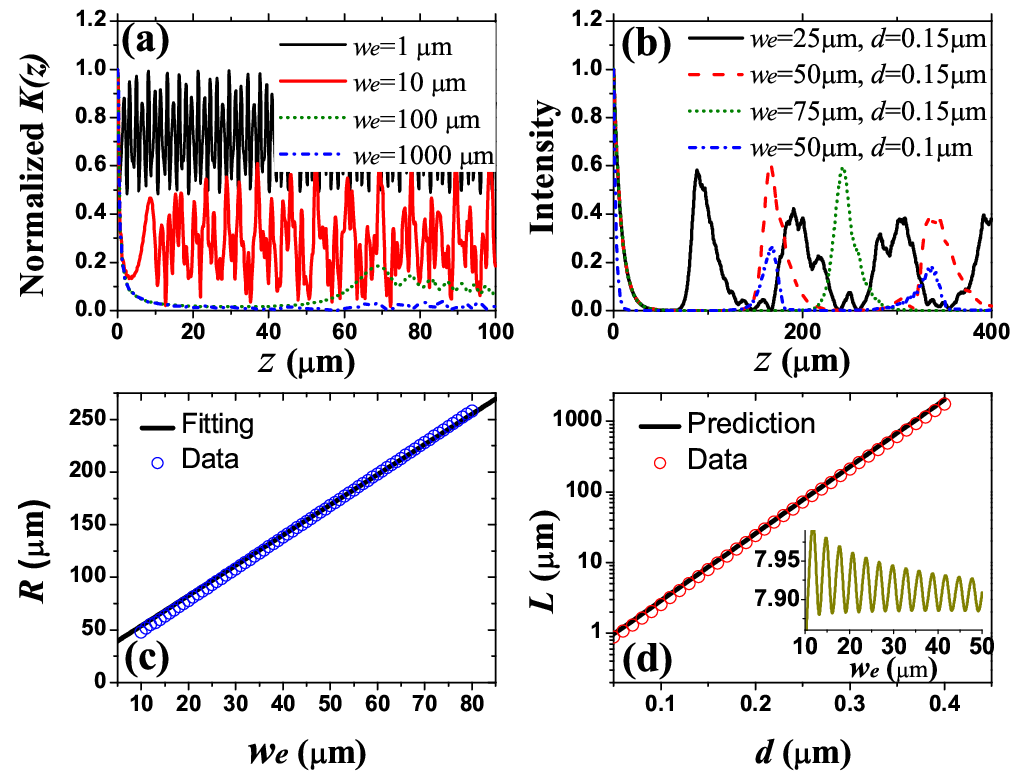} \\
\caption{(Color online) (a) The memory kernel function for different environment
sizes, with $d=0.15\mu m$. (b) The dynamics of system for different
$w_{e}$ and $d$. (c) The revival period $R$ against $w_{e}$, with
dots obtained from the system dynamics, and solid line from the analytical
model, with $d=0.15\mu m.$ (d) The decay length against $d$, with
dots and line being results of the system dynamics and analytical
model with $w_{e}=\infty$. Inset: the decay length against $w_{e}$
obtained from the system dynamics.}

\label{f3}
\end{figure}

A direct view of the dynamics of the coupled system-environment can
be clearly observed in Fig. 1(c)-(e). When $w_{e}$ is comparable with $\lambda$,
there is few modes can be involved to interaction with the system,
the coherent coupling of which leads to sinusoidal oscillation of
system energy {[}Fig. 1(c){]}. For larger $w_{e}$, more modes could
interact with the system, giving rise to a complex dynamics. It is
interesting that the corporation of environment modes shows a classical
ray trajectory of light leaked from the system waveguide, as indicated
by the dashed line in Fig. 1(d), with an incident angle $\chi=\mathrm{arcsin}(\beta_{0}/k)$
with respect to $z$-axis. For $w_{e}\geq10\mu m$, the energy exponentially
decays with $z$ and can almost completely leak into environment {[}Fig.
1(e){]}. And after a certain distance, the leakage will be reflected
into system and cause a revival of system. 

Furthermore, some important interaction properties to understand the underlying physics can be easily obtained from those figures. The formal solution of the dynamics
of system $a(z)$ is derived in the interaction picture $H_{I}(z)=\sum_{j}\hbar g_{j}(a^{\dag}b_{j}e^{i(\beta_{0}-\beta_{j})z}+ab_{j}^{\dag}e^{-i(\beta_{0}-\beta_{j})z})$
by employing the formal solution $b_{j}(z)=-ig_{j}\int_{0}^{z}a(\tau)e^{-i(\beta_{0}-\beta_{j})\tau}d\tau$
. The dynamics of system reads
\begin{equation}
\frac{\partial}{\partial z}a(z)=-\int_{0}^{z}a(\tau)K(\tau-z)d\tau,
\end{equation}
with the memory kernel function $K(\tau)=\sum_{j}g_{j}^{2}e^{-i(\beta_{0}-\beta_{j})\tau}$
\citep{Breuer}. For small $w_{e}$, the beating of multiple environment
modes gives rise to the oscillation behavior of $K(\tau)$. The non-zero
memory effect leads to non-Markovian dynamics of system, showing non-exponential
energy decay and revival phenomenon {[}as shown in Fig. 1(b){]}. Since
the period of $K(\tau)$ is determined by the environment mode density
$\rho(n)\approx\frac{kw_{e}}{\pi}\frac{n}{\sqrt{n_{d}^{2}-n{}^{2}}}$,
the non-Markovian revival period linearly depends on $w_{e}$, which
can be observed in Fig. 2(b). We can also deduce the revival period $R=2w_{e}\mathrm{tan}\chi+R_{0}$
with intuitive understanding of revival as beam reflection at boundary
of environment. The revival period extracted from the dynamics of
system shown in Fig. 2(c) are well fitted by this formula. The constant
$R_{0}$ is corresponding to an extra distance required for tunneling
\citep{hauge1989tunneling} and Goos-Hänchen shift \citep{snyder1976goos}
of light when reflecting at the environment boundary.

Fig. 2(a) shows that $K(\tau)$ becomes the Dirac delta function $\delta(\tau)$
when the environment size approaches to infinity. As a great many
environment modes are involved when $w_{e}\rightarrow\infty$, the
memory kernel function can be approximately written in the form of
integration $K(\tau)=\int g^{2}(n)e^{-i(\beta_{0}-nk)\tau}\rho(n)dn$,
where the coupling strength $g(n)$ is a function of mode index $n$
{[}Fig. 1(a){]}. Then
\begin{equation}
K(\tau)\approx J(n_{0})\delta(\tau),\label{NSD}
\end{equation}
where $n_{0}=\beta_{0}/k$ and $J(n)=g^{2}(n)w_{e}\frac{n}{\sqrt{n_{d}^{2}-n^{2}}}$
is the spectrum density function of the reservoir. This $\delta(\tau)$
memory function gives rise to the Markovian process that system energy
decays exponentially as $a(z)=e^{-z/2L}$. The decay length
\begin{equation}
L=\frac{1}{2J(n_{0})}\approx L_{0}e^{-2\sqrt{\beta_{0}^{2}-k^{2}}d},
\end{equation}
where $L_{0}=\frac{1}{k}\frac{n_{0}(n_{d}^{2}-1)^{2}(kw_{s}+2/\sqrt{n_{0}^{2}-1})}{8(n_{0}^{2}-1)(n_{d}^{2}-n_{0}^{2})^{3/2}}$
is solved explicitly. In Fig. 2(d), the analytical decay length is
compared to the results extracted from the dynamics of system with
logarithmic fitting. Our prediction is consistent with the data, and
the slight discrepancy is due to non-flat noise spectrum density {[}Eq.
(\ref{NSD}){]}. It is surprising that the formula of $L$ in the
continuum limit works for small $w_{e}$, as shown in the inset of
Fig. 2(d). The oscillation of which is due to the variation of mode
index of discrete modes in finite environment.

\begin{figure}
\includegraphics[width=8cm]{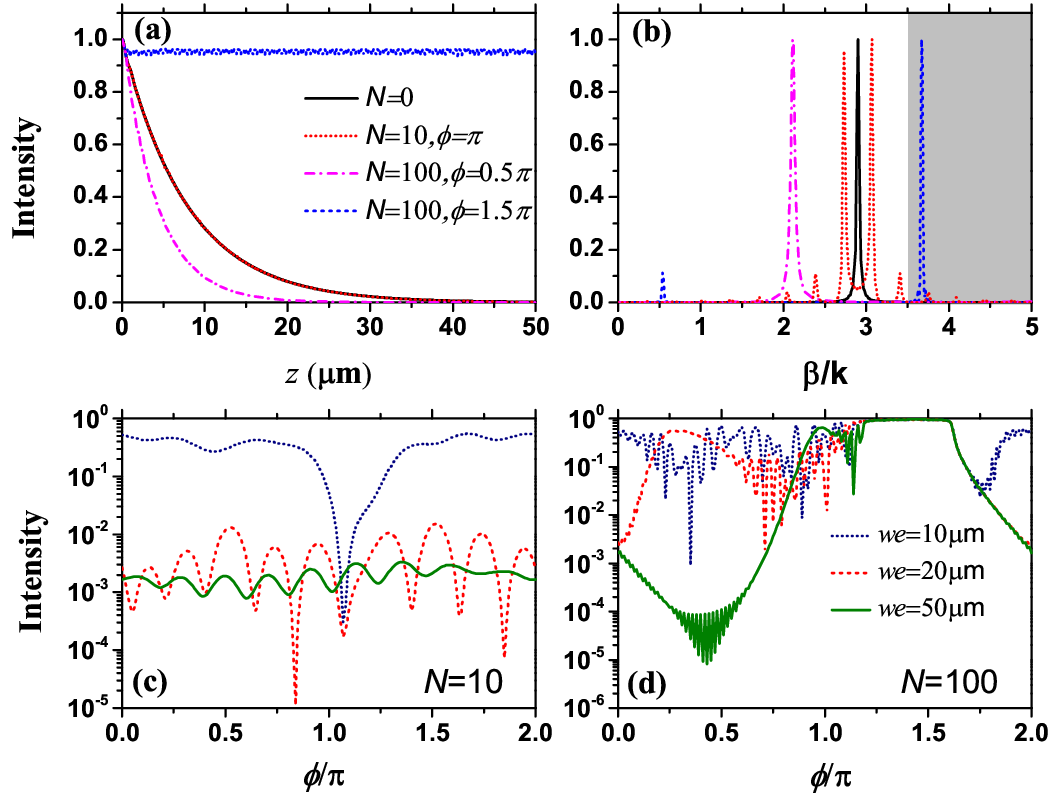} \\
 \caption{(Color online) The system evolution (a) and corresponding spectrum
(b) when a sequence of $N$ modulation of phase $\phi$ applied to
the system. (c) and (d) The system energy at $z=50\mu m$ against
$\phi$ for different $w_{e}$ and $N$, with $d=0.15\mu m$. }

\label{f4}
\end{figure}

\emph{Dynamical Decoupling.-} It is well known that external modulation
to open quantum systems can modify the energy decay and decoherence,
and the so-called ``dynamical decoupling'' technique has been widely
adopted to keep the coherence of electron spins \citep{PhysRevLett.82.2417,du2009preserving,de2010universal}.
Similarly, for the photonic simulation of system-environment interaction,
a dynamical modulation of system waveguide can also alter the energy
decay. Here, a sequence of $N$ modulations with equal interval is
applied to the system waveguide, where each modulation corresponds
to an abrupt change of phase $\phi$. Fig. 3(a) shows the evolution
of system changes when modulation is applied. The dissipation of system
can be enhanced or inhibited significantly depending on $\phi$ and
$N$. As shown by Fig. 3(c) and (d) are energy in the system waveguide
at $z=50\mu m$ against the modulation phase $\phi$, with different
$w_{e}$ and $N$.

There's an intuitive way to understand the modified decay: The modulations
add extra phase to the propagating light, which is equivalent to an
increase or decrease of the effective index $\widetilde{n}$, as the
phase of propagating photon is proportional to $\widetilde{n}$. As
shown in Fig. 3(b), the spectrum of the system is shifted by the modulations.
According to the spectrum density of reservoir {[}Fig. 1(a){]}, different
$\widetilde{n}$ corresponds to different coupling strength, then the
modulations give rise to acceleration or deceleration of dissipation.
When $\widetilde{n}$ is larger than the cut-off index $n_{d}$, the
dissipation of energy in system to the environment is forbidden. From
another point of view, the waveguide with periodic modulation is similar
to the photonic crystal or grating structures, which will induce a
band structure to the light. By this means, noise in environment can
be prevented from propagating in system. These provide physics insight
to the dynamical decoupling in time dimension where a sequence of
modulation in time-axis is applied: the effective frequency of the
system shifts to a higher value than the cut-off frequency of bath.
Or, we can image a crystal or band structure in time dimension, where
only the signal with certain frequencies can enter and be kept in
the system while most of the broadband noises in environment is blocked.

\begin{figure}
\includegraphics[width=8cm]{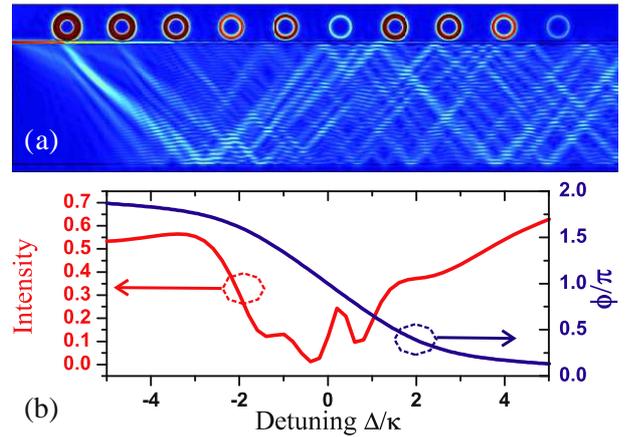} \\
 \caption{(Color online) (a) The fields of system and environment when microdisks
are added to the waveguide to induce a phase modulation. (b) The dynamics
against the detuning of input light to the whispering gallery modes.}

\label{f5}
\end{figure}

One possible way to realize such phase modulation sequence is shown
in Fig. 4(a), with whispering gallery mode (WGM) microdisks coupling
to the system waveguide \citep{cai2000observation}. The microdisk
with radius $r=1\mu m$ has intrinsic quality factor higher than $10^{6}$.
When put close to the system waveguide with a gap ($0.15\mu m$),
its loaded quality factor is only about $2000$. That means the WGM
is working in the strongly over coupling regime with $\kappa_{e}\gg\kappa_{i}$,
where $\kappa_{i(e)}$ is the intrinsic (external) loss. When the
light in system passes the over-coupled WGM resonator, the change
of transmitted light is $T(\Delta)=-\frac{1-i\Delta/(\kappa_{e}-\kappa_{i})}{1+i\Delta/(\kappa_{e}+\kappa_{i})}\approx-\frac{1-i\Delta/\kappa_{e}}{1+i\Delta/\kappa_{e}}$,
where $\Delta$ is the frequency detuning to the resonance. As $|T(\Delta)|\approx1$,
the system acquires a modulation of phase $\phi(\Delta)=\mathrm{arg}[T(\Delta)]$.
Numerical simulation of the dynamical decoupling with $N=10$ for
$w_{e}=10\mu m$ is performed. Comparing the mode profile of Fig.
4(a) with Fig. 1(e), the evolutions are significantly changed by modulations.
From Fig. 4(b), a strong modification of the decay is shown around
the resonance that phase modulation $\phi\approx\pi$, which is consistent
with the prediction in Fig. 3(a).

\emph{Conclusion.-} The Markovian, non-Markovian processes and dynamic
decoupling of open quantum systems are studied in a photonic simulation
of system-environment interaction. Intuitive physical insight and
deep understanding of these phenomena can be gained from the direct
view of electromagnetic field profiles. Our study also opens the possibility
to tune the dissipation in photonic system, similar to the dynamic
decoupling of spins.

\textbf{Acknowledgments} This work was supported by the 973 Programs (No.2011CB921200 and No.
2011CBA00200), the National Natural Science Foundation of China (NSFC) (No.
11004184), the Knowledge Innovation Project of the Chinese Academy of
Sciences (CAS).


\end{document}